\documentclass[conference]{IEEEtran}
\IEEEoverridecommandlockouts
\usepackage{cite}
\usepackage{amsmath,amssymb,amsfonts}
\usepackage{algorithmic}
\usepackage{graphicx}
\usepackage{textcomp}
\usepackage{xcolor}
\begin{document}

\title{Link~Streams~as~a~Generalization~of Graphs~and~Time~Series}

\author{
\IEEEauthorblockN{Esteban Bautista}
\IEEEauthorblockA{\textit{IMT Atlantique} \\
\textit{Lab-STICC, CNRS UMR 6285}\\
F-29238 Brest, France\\
esteban.bautista-ruiz@imt-atlantique.fr}
\and
\IEEEauthorblockN{Matthieu Latapy}
\IEEEauthorblockA{\textit{Sorbonne Université} \\
\textit{ LIP6, CNRS UMR 7606} \\
F-75005 Paris, France \\
matthieu.latapy@lip6.fr}
}

\maketitle
\thispagestyle{plain}
\pagestyle{plain}

\begin{abstract}
A link stream is a set of possibly weighted triplets $(t,u,v)$ modeling that $u$ and $v$ interacted at time $t$. Link streams offer an effective model for datasets containing both temporal and relational information, making their proper analysis crucial in many applications. They are commonly regarded as sequences of graphs or collections of time series. Yet, a recent seminal work demonstrated that link streams are more general objects of which graphs are only particular cases. It therefore started the construction of a dedicated formalism for link streams by extending graph theory. In this work, we contribute to the development of this formalism by showing that link streams also generalize time series. In particular, we show that a link stream corresponds to a time-series extended to a relational dimension, which opens the door to also extend the framework of signal processing to link streams. We therefore develop extensions of numerous signal concepts to link streams: from elementary ones like energy, correlation, and differentiation, to more advanced ones like Fourier transform and filters. 
\end{abstract} 


\section{Introduction}

Numerous systems generate data as time-stamped interactions between individuals. Such time-stamped interactions may represent a computer transmitting a packet to another computer at a given time, a person calling another person at a given time, or a bank account transferring money to another bank account at a given time, to list a few examples. Such data can be very well modeled as a link stream \cite{zhou2020data, cazabet2019challenges, citraro2022delta, wilmet2018degree, boudebza2018olcpm}: a set of possibly weighted triplets $(t,u,v)$ indicating that $u$ and $v$ interacted at time $t$. The proper study of link streams is crucial for a better understanding of the systems generating the data and for the timely identification of potentially harmful events, like network attacks or monetary frauds  \cite{fontugne2017scaling, fontugne2016characterizing, bhatia2020midas, chang2021f, eswaran2018sedanspot}. As a result, the development of techniques allowing a precise understanding of the properties and events contained withins a link stream is a subject of utmost importance.

Link streams are traditionally studied as collections of graphs (one per time-stamp $t$) or time series (one per pair $u,v$). These perspectives allow to readily employ graph or time-series methods to study them, however at the cost of not allowing to study their structure and dynamics jointly. The work of \cite{latapy2018stream} proposed to address this issue through a change of perspective: a link stream should not be considered a collection of graphs or time series but rather an individual mathematical object of study. Indeed, \cite{latapy2018stream} shows that, rather than being a graph sequence, a link stream corresponds to a more general entity of which graphs are only arise as a particular case: a graph is a link stream with no dynamics. Based on this observation, \cite{latapy2018stream} and subsequent works \cite{Latapy2019,simard2019computing, baudin2023lscpm, ghanem2017ego, naima2023temporal, VIARD2016245} have extended numerous concepts from graph theory to link streams in order to lay the foundations of a dedicated formalism which aims to be for link streams what graph theory is for graphs. The formalism includes link stream versions of concepts such as density, neighborhood, cliques, clustering coefficient, paths, etc. The characterizing property of these extensions is that they all reduce to their classical graph version when the link stream has no dynamics \cite{latapy2018stream}. 

In this work, we complement this vision by showing that link streams also generalize time series. In particular, we show that a link stream can be seen as a time series extended to a relational dimension, implying that classical time-series correspond to link streams lacking structure. The main difference between our approach and the one of \cite{latapy2018stream} is that the latter represents link streams as a set of time-stamped edges, while we represent them as an indicator (or weight) function supported on time-stamped relations. These two representations encode for the same information but admit different mathematical analyses. In particular, by representing a link stream through a function, we open the door to do signal processing directly on them. This is, we bring the possibility to measure their variability, to compare link streams, to decompose a complex link stream into a combination of simpler ones, or to filter-out undesired information from them, to list some cases. These concepts exploit the vector space nature of the signals and therefore are not easy to derive from the set-theoretical perspective employed in \cite{latapy2018stream}. The aim of this work is to develop extensions of the core concepts from signal processing to link streams. 

The paper is structured as follows. Section \ref{Sec.graphs} briefly reviews the perspective of link streams as a generalization of graphs. In particular, it presents the core link stream formalism that has been developed by extending graph theory. Then, Section \ref{Sec.signals} addresses their vision as a generalization of time-series. The section develops extensions of some of the most used concepts in signal processing: from simple ones like product, shifting and differentiation, to more advanced ones like decomposition and filtering. Lastly, Section \ref{Sec.conclusion} concludes the work. 
\section{Link streams as a generalization of graphs}\label{Sec.graphs}

We now briefly review the formalism for link streams that has been constructed by generalizing graph theory. We mainly focus on the developments of \cite{latapy2018stream}, albeit we recall that newer works continue to enrich the formalism \cite{Latapy2019,simard2019computing, baudin2023lscpm, ghanem2017ego, naima2023temporal, VIARD2016245}. Throughout this section, we consider continuous-time undirected and unweighted link streams, which are defined as the triplet $L = (T, V, E)$, where $V$ denotes a set of nodes of size $n = |V|$, $T$ is an interval of $\mathbb{R}^+$ representing a set of times, and $E \subseteq T \times V \otimes V$ corresponds to a set of time-stamped undirected edges\,\footnote{$V \otimes V$ denotes the unordered pairs of distinct elements of $V$, that we denote by $uv \in V\otimes V$.}. As mentioned above, link streams are considered to generalize graphs. This relationship is established as follows. Let $G(L) = (V, E(L))$ be the graph induced by $L$, where $E(L) = \{uv : (t, uv) \in E \}$. Then, if $E = T \times E(L)$, we have that $L$ is called a graph-equivalent stream. A graph-equivalent stream is therefore a link stream in which edges have no dynamics, meaning that it can be entirely recovered from its induced graph, which aggregates all links that appear at some point. This is the crucial equivalence that the formalism of \cite{latapy2018stream} aims to exploit: it aims to propose extensions of graph concepts to link streams so that the concept reduces to the graph one when it is applied to a graph-equivalent stream. 

One of the most elementary properties of a graph is its number of edges. In link streams, it is not as trivial to determine their number of edges due to the continuous nature of the time domain. Thus, it is crucial to begin the extension of graph theory from such fundamental concepts, as more elaborate ones depend on them. The definition proposed by \cite{latapy2018stream} is that the number of links of $L$ is given as:
\begin{equation}\label{eq.num_edges}
m = \frac{|E|}{|T|}.
\end{equation}
The interpretation of (\ref{eq.num_edges}) is that each pair $uv$ contributes to the number of links proportionally to the time it is active. This is a sound extension as, in a graph-equivalent stream, it holds that $m = \frac{|T \times E(L)|}{|T|} =  |E(L)|$, which is the classical definition for graphs.

In graphs, a concept that is tightly related to the number of edges is the one of density. The density of a graph can be interpreted as the probability that, when a pair $uv$ is selected at random, the pair exists in the edge-set of the graph. The density of a link stream is therefore defined as the probability of randomly selecting a triplet $(t, uv) \in T \times V \otimes V$ and that the triplet is included in $E$. This is expressed as:
\begin{equation}\label{eq.graph_density}
\delta(L) = \frac{|E|}{|T \times V \otimes V |}
\end{equation} 
Since $|V \otimes V | = \frac{n(n-1)}{2}$, we have that (\ref{eq.graph_density}) can be written as $\delta(L) = \frac{2 m}{n(n-1)}$. Therefore, it can be clearly seen that for a graph-equivalent stream (\ref{eq.graph_density}) reduces to the classical definition of density for undirected unweighted graphs.

In graphs, the concept of density is useful to define certain sub-structures, like cliques. This is extended to link streams as follows. Let $T' \subseteq T$, $V' \subseteq V$ and $C = T' \times V'$ denote a cluster of temporal nodes. Moreover, let $L' = (T', V', E \cap \{T' \times V' \otimes V'\} )$ denote the sub-link stream of $L$ induced by the restrictions $T'$ and $V'$. Then, we say that the cluster $C$ is a clique of $L$ if $\delta(L') = 1$. Notice that if $L$ is a graph-equivalent stream and $T' \times V'$ is a clique of $L$, then $V'$ is necessarily a clique in the graph induced by $L$. 

In a graph, the degree of a node depends on the size of its neighborhood. In link streams, it is less clear to define the degree of nodes as the concept of neighborhood must now take into account the temporal dimension. In particular, \cite{latapy2018stream} defines the neighborhood of a node $v \in V$ as the following cluster: 
\begin{equation}
N(v) = \{ (t,u) : (t, uv) \in E \} 
\end{equation}
The degree of $v$ is thus defined as the number of links it has with its neighbors:
\begin{equation}
d(v) = \frac{|N(v)|}{|T|}
\end{equation} 
Similarly, it is natural to define instantaneous neighborhood and degree by letting $N_t(v) = \{u : (t, uv) \in E\}$ be the instantaneous neighborhood of $v$ at time $t$ and $d_t(v) = |N_t(v)|$ denote the instantaneous degree of $v$ at time $t$. Observe that if $L$ is a graph-equivalent stream, then the neighborhood of $v$ can be expressed as $N(v) = T \times V'$ and $V'$ is indeed the neighborhood of $v$ in the graph induced by $L$. Moreover, in a graph-equivalent stream the degree reduces to $d(v) = |V'|$, which is the same degree that $v$ has in the induced graph. The same argumentation applies for the instantaneous degrees. 

In a graph, the clustering coefficient of a node indicates the probability that two of its neighbors chosen at random are connected. In a link stream, the clustering coefficient of a node $v$ is therefore defined as the probability that two of its neighbors chosen at random and at the same time are connected. To formalize this, let $E_t(N(v)) = \{ vw : (t, uw) \in E, (t,u) \in N(v), (t,w) \in N(v) \}$ be the edges between the neighbors of $v$ at time $t$. The clustering coefficient of $v$ is thus defined as:
\begin{equation}
cc(v) = \frac{ \int_{t\in T} |E_t| dt}{ \int_{t \in T} |N_t(v) \otimes N_t(v)|}
\end{equation} 
Notice that in a graph-equivalent stream $E_t$ and $N_t(v)$ are constant over time, thus they go out of the integrals making $cc(v)$ coincide with the clustering coefficient of the graph induced by $L$.  

A closely related notion in graphs is the $k$-core, which refers to the largest cluster of nodes satisfying the property that all the vertices in the subgraph induced by the cluster have a degree no less than $k$. The concept of $k$-core extends to a link stream $L$ as the largest cluster $C^k = T' \times V'$, where $T' \subseteq T$ and $V' \subseteq V$, such that, for all $(t,v) \in C^k$, $d_t(v) \geq k$ in $L' = (T', V', E \cap \{T' \times V' \otimes V'\} )$, where $L'$ is the sub-link stream of $L$ induced by the restrictions $T'$ and $V'$. If $C^k = T' \times V'$ is the $k$-core in a graph-equivalent stream $L$, then we have that $V'$ is the $k$-core in the graph induced by $L$ as $d_t(v)$ is independent of time.

In graphs, a path refers to an sequence of edges that respect the adjacency of vertices. It is an important concept upon which numerous others depend, such as connectedness, diameter, cycles, random walks, centralities, etc. In link streams the extension is not trivial, as there are two degrees of freedom to define the order: by respecting the order of the time axis or by respecting the adjacency of vertices. In \cite{latapy2018stream}, paths in link stream $L = (V, T, E)$ are defined as follows. A path $P$ from $(\alpha, u) \in T \times V$ to $(\omega, v) \in T \times V$ is a sequence $(t_0, u_0, v_0), (t_1, u_1, v_1), \dots, (t_k, u_k, v_k)$ of elements in $E$ such that $u_0 = u$, $v_k = v$, $t_0 \geq \alpha$, $t_k \leq \omega$, $t_i \leq t_{i+1}$ for all $i$, and $v_i = u_{i+1}$. From this definition, then paths in link streams have both a length $k + 1$ (the number of triplets in the path) and a duration $t_k - t_0$ (the time delay from start to finish). This makes paths in link streams radically different to paths in graphs which only have the notion of length. The following definitions formalize these differences. 

Consider a path $P$ from $(\alpha, u)$ to $(\omega, v)$. Then, $P$ is said to be a shortest path if it has minimal length and the length of the shortest path between $(\alpha, u)$ to $(\omega, v)$ defines their distance. Additionally, $P$ is said to be a fastest path if it has minimal duration and the duration of the fastest path between $(\alpha, u)$ to $(\omega, v)$ defines their latency. If $\omega \leq t$ is the minimal value of time such that there is a path from $(\alpha, u)$ to $(\omega, v)$ then $P$ is said to be a foremost path between $(\alpha, u)$ to $(t, v)$ and $\omega - \alpha$ defines the time to reach $v$ from $(\alpha, u)$. Lastly, a link stream $L$ is said to be strongly connected if for all $(\alpha, u) \in T \times V$ and $(\omega, v) \in T \times V$ there is a path from $(\alpha, u)$ to $(\omega, v)$ in $L$. Observe that in a graph-equivalent stream $L$, all the reachable vertices can be attained with zero latency. Hence, this implies that there is a path between two vertices in $L$ if an only if such path exists in the graph induced by $L$. This also implies that shortest paths in $L$ have the same length as shortest paths in its induced graph. 
\section{Link Streams as a generalization of signals}\label{Sec.signals}

In this section, we enrich the toolkit of Section \ref{Sec.graphs} by showing that link streams also generalize time series. In particular, we leverage such insight to extend signal processing concepts to link streams. We highlight that some of these extensions have already been presented in \cite{Bautista_ls_2023} for the discrete time case. Here, we target the continuous time case. Therefore, this section considers the following extended definition of a link stream: a quintuplet $L = (T, V, R, \mathcal{L}, \mathcal{I})$, where $T$ and $V$ are defined as above, $R \subseteq V \times V$ is an arbitrary subset of $M$ directed relations, $\mathcal{L}: T \times R \to \mathbb{R}$ is a function assigning a weight to time-stamped relations, and $\mathcal{I}:R \to [0, M-1]$ is a function that orders the elements of $R$. Thus, there are two essential differences between our definition and the used in Section \ref{Sec.graphs}: (i) we do not model temporal interactions through the set $E$ but rather through the function $\mathcal{L}$; and (ii) we incorporate $R$ and $\mathcal{I}$ as means to have control over the domain of $\mathcal{L}$. For the sake of notation lightness, throughout this section we refer to a link stream $L = (V, T, R, \mathcal{L}, \mathcal{I})$ through its weight function $\mathcal{L}$ unless there is need to specify the other elements of the quintuplet.  

The main advantage of representing $L$ through $\mathcal{L}$ rather than $E$ is that $\mathcal{L}$ lives in a vector space. In this work, we assume this space to be restricted to all square integrable functions supported on $T \times R$: this is, of all link streams whose sum of squared weights is finite (essentially all link streams that appear in practice) and that share $T$ and $R$ in their definitions. The fact that link streams form a vector space means that we can perform algebraic operations to and between them: scalings, addition, products, expansions, approximations, change of basis, etc. For instance, we may take two link streams and add/multiply them in order to produce a new one; or we may define a set of link streams that form a basis and use them to form any other link stream defined on the same set of times and relations. These are concepts that do not naturally arise from the perspective of graphs used in Section \ref{Sec.graphs}, yet they naturally do from the perspective of signal processing which meaningfully defines them for time series. Notably, $\mathcal{L}$ corresponds to a time series extended to a relational dimension: from $\mathcal{L} : T \to \mathbb{R}$ being a classical time series to $\mathcal{L} : T \times R \to \mathbb{R}$ being a link stream. Thus, classical time series are indeed link streams with no structure. This motivate us to extend signal processing concepts to link streams by making them take into account the new dimension. As a design principle, we aim that a proposed extension reduces to the classical definition for time series when it is applied to a link stream with no structure. 

It is important to mention that the signal processing framework contains numerous concepts that exploit the ordered nature of the time dimension. Yet, in a link stream, the relational dimension has no natural order and extending such concepts to an unordered domain is not very significant. The introduction of $R$ and $\mathcal{I}$ address this situation. Particularly, $\mathcal{I}$ allows us to set an order to relations and $R$ allows us to choose the structure in which it is meaningful to apply such order. For example, we may choose $R$ to be the outgoing relations of a node and $\mathcal{I}$ to rank such relations according to their probability of observation. Another example may be that $R$ is chosen to represent the relations within a community and that $\mathcal{I}$ ranks them according to their importance. It can also be that $R$ denotes a path and $\mathcal{I}$ captures the order in which relations appear in the path. During the section, we discuss various choices of $R$ and $\mathcal{I}$ that may be interesting in a given application scenario. 

\subsection{Extension of elementary concepts}
We begin our extension by stressing that signal processing concepts can be readily applied to a link stream by using them relation-wise. Namely, suppose that $o$ refers to a signal processing concept that modifies a time-series $x$ into a new one $x'$ through some transformation of the form $x'(t) = \int c(\tau) x(t -\tau) d \tau$, where $t \in T$. We have that $o$ can be readily applied to a link stream $\mathcal{L}$ by fixing $e \in R$ and letting $t$ run as $\mathcal{L}'(t, e)  =  \int c(\tau) \mathcal{L}(t -\tau, e) d \tau$, which is done for all $e \in R$. Therefore, to extend $o$ to the relational dimension two options arise. The first one is to define a relational analog of $o$, called $g$, that is now applied by fixing $t$ and letting $e$ run, which is done for all $t$. The link stream extension thus consists in the combined application of $o$ and $g$: the former addressing the temporal dimension and the latter the relational one. The second option consists in revamping the definition of $o$ to consider both variables $t$ and $e$ jointly. In this work, we address the former, which allows us to focus in finding meaningful ways to leverage the relational information while letting raw signal processing concepts handle the temporal one. We leave the extension of concepts based on the second approach as future work.

{\bf Scaling, addition and product.}~Most signal processing concepts rely on the fact that signals can be scaled, added and multiplied. We now formalize these operations for link streams. Let $\mathcal{L}$ denote a link stream, we define its scaling by a constant $c$ as the link stream $(c\mathcal{L})(t,e) = c\mathcal{L}(t,e)$. Let $\mathcal{L}_1$ and $\mathcal{L}_2$ be link streams defined on the same $T$ and $R$, we define their addition as the link stream $(\mathcal{L}_1 + \mathcal{L}_2)(t,e) = \mathcal{L}_1(t,e) + \mathcal{L}_2(t,e)$. Let $\mathcal{L}_1$ and $\mathcal{L}_2$ be link streams defined on the same $T$ and $R$, we define their product as the link stream $(\mathcal{L}_1 \cdot \mathcal{L}_2)(t,e) = \mathcal{L}_1(t,e) \cdot \mathcal{L}_2(t,e)$. These definitions trivially follow from the fact that link streams sharing $T$ and $R$ form a vector space. Moreover, it can be straightforwardly seen that they reduce to the classical time-series definition when $R$ is a singleton. 

{\bf Correlation, energy, distance.}~These concepts are commonly employed to characterize and compare time series. We now define them for link streams. Let $\mathcal{L}_1$ and $\mathcal{L}_2$ be link streams defined on the same $T$ and $R$, we define their zero-lag correlation as the scalar: 
\begin{equation}\label{eq.correlation}
\text{corr}( \mathcal{L}_1, \mathcal{L}_2) = \sum_{e \in R} \int_{t \in T} (\mathcal{L}_1 \cdot \mathcal{L}_2)(e, t) dt
\end{equation}
In signal processing, the zero-lag correlation is a measure of the similarity between two signals. Notice that (\ref{eq.correlation}) admits the same interpretation for link streams and that, in the case $\mathcal{L}_1$ and $\mathcal{L}_2$ are unweighted, the following interesting property holds: $\text{corr}( \mathcal{L}_1, \mathcal{L}_2) = 2 |E_1 \cap E_2 |$, where $E_1$, $E_2$ are the set equivalents of $\mathcal{L}_1$ and $\mathcal{L}_1$ used in Section \ref{Sec.graphs}. Moreover, we have that when $R$ is a singleton, (\ref{eq.correlation}) reduces to the standard definition of zero-lag correlation for time series. 

In signal processing, the energy of a signal is the zero-lag correlation of a signal with itself. Thus, given a link stream $\mathcal{L}$, we define its energy as:
\begin{equation}\label{eq.energy}
\text{energy}(\mathcal{L}) = \text{corr}( \mathcal{L}, \mathcal{L}) 
\end{equation}
Interestingly, the following property holds when $\mathcal{L}$ is unweighted: $\text{energy}( \mathcal{L} ) = 2|E| $. This indicates that the energy of a link stream is a measure of the amount of information contained in it, which is the same interpretation that energy has for time series. Moreover, this property shows that the energy of an unweighted link stream is $2|T|$ times its number of links $m$ (see Equation (\ref{eq.num_edges})). Thus, this opens the door to use (\ref{eq.energy}) as a means to extend the definition of $m$ to weighted link streams.

In signal processing, the distance between two time series is measured as the square root of the energy of their difference. Thus, given two link streams $\mathcal{L}_1$ and $\mathcal{L}_2$ defined on the same $T$ and $R$, we define their distance as
\begin{equation}\label{eq.distance}
d( \mathcal{L}_1, \mathcal{L}_2 ) = \text{energy}(\mathcal{L}_1 - \mathcal{L}_2)^{1/2}
\end{equation}
Interestingly, when $\mathcal{L}_1$ and $\mathcal{L}_2$ are unweighted, the following property holds:  $d( \mathcal{L}_1, \mathcal{L}_2 ) = \sqrt{2 |E_1 \triangle E_2 |}$, where $\triangle$ denotes to the geometric difference operator. Thus, the distance of two unweighted link streams is strongly related to their edit distance (number of different links). This opens the door to consider (\ref{eq.distance}) as an extension of the edit distance in a weighted scenario. 

{\bf Regularity, shift, cross-correlation.}~These are basic concepts in signal processing that exploit the ordered nature of the time domain. Therefore, we extend them to link streams by leveraging the order of $R$ provided by $\mathcal{I}$. In signal processing, the Lipschitz regularity constant of a time series $x$ gives an indication of how much $x$ varies. Formally, it is defined as the smallest constant $r$ for which the following inequality holds: $|x(t_1) - x(t_2)| \leq r |t_1 - t_2|$ for all $t_1, t_2 \in \mathbb{R}$. We extend this concept to link streams as follows. For a link stream $\mathcal{L}$, we define its $T$-based Lipschitz regularity constant at point $(t_i,e_k) \in T \times R$, where $k$ is the index associated to $e$ by $\mathcal{I}$, as the minimum constant $r_T(t_i, e_k)$ for which the following inequality holds:
\begin{equation}\label{eq.T_regularity}
|\mathcal{L}(t_i, e_k) - \mathcal{L}(t_j, e_k)| \leq r_T(t_j,e_k) |t_i - t_j| ~~\forall t_j \in T 
\end{equation}
Similarly, we define the $R$-based Lipschitz regularity constant of $\mathcal{L}$ at point $(t_i,e_k)$, as the minimum constant $r_R(t_i, e_k)$ for which the following inequality holds:
\begin{equation}\label{eq.R_regularity}
|\mathcal{L}(t_i, e_k) - \mathcal{L}(t_i, e_\ell)| \leq r_R(t_i,e_k) |k-\ell| ~~\forall e_\ell \in R
\end{equation}
Then, we define the total Lipschitz regularity constant of $\mathcal{L}$ at point $(t_i,e_k)$ as:
\begin{equation}\label{eq.total_regularity}
r(t_i, e_k) = \max \{r_T(t_i,e_k), r_R(t_i,e_k) \}
\end{equation}
Lastly, we define the Lipschitz regularity constant of $\mathcal{L}$ as:
\begin{equation}\label{eq.ls_regularity}
\text{reg}(\mathcal{L}) = \max_{(t,e_k) \in T \times R} r(t,e_k)
\end{equation}
As it can be seen, $r_T(t_i,e_k)$ measures variations of $\mathcal{L}(t_i,e_k)$ across $T$, while $r_R(t_i,e_k)$ does it across $R$. Interestingly, by properly choosing $\mathcal{I}$, $r_R(t_i,e_k)$ can provide as equally meaningful insights about $\mathcal{L}(t_i,e_k)$ as $r_T(t_i,e_k)$. For example, suppose that $R$ represents the internet backbone links and that $\mathcal{I}$ indexes the elements of $R$ in decreasing order of their bandwidth capacity. Thus, if the link stream weights represent the amount of traffic flowing through those links over time, then we have that a large value of $r_R(t_i,e_k)$ indicates that $e_k$ at time $t$ is being under or over utilized with respect to other links having similar bandwidth. This is a type of variability not measured by $r_T(t_i,e_k)$ which, on the other hand, would allow to identify if the traffic flowing through $e_k$ is stable over time. Hence, $r_R(t_i,e_k)$ and $r_T(t_i,e_k)$ supplement each other. Additionally, notice that $r_R(t_i,e_k) = 0$ when $R$ is a singleton, showing that our extension reduces to the time series one when the link stream has no structure. 

In signal processing, shifting refers to the movement of a time series in time. For a time series $x$ and a shift amount of $\tau \in \mathbb{R}$ units, $x'(t) = x(t - \tau)$ defines the shifting of $x$ by $\tau$. Therefore, we extend shifting to link streams by moving them across both the time and relational axes. We stress that for a link stream defined on a bounded interval $T$, moving it in time may send it outside the interval for which it is defined. Thus, when dealing when shifts, we assume the temporal domain of the concerned link stream to be extended to $\mathbb{R}$ and that it is filled with zeros on such extended region. Namely, if $\mathcal{L}$ is originally defined on $T \times R$, we assume it is extended to $\mathbb{R} \times R$ such that $\mathcal{L}(t, e) = 0$ for all $t \in \mathbb{R} \setminus T$. Based on this extended link stream, we define shifting as follows. Given a link stream $\mathcal{L}$ and shift amounts $\tau \in \mathbb{R}$ and $\eta \in \mathbb{Z}$, we define the shifting of $\mathcal{L}$ by $\tau$ and $\eta$ as the link stream 
\begin{equation}\label{eq.shift}
\mathcal{S}_{\tau, \eta}( \mathcal{L} )(t_i, e_k) = \mathcal{L}(t_i - \tau, e_{k'})
\end{equation}
where $k' = |k - \eta| \mod M$ if $k \geq \eta$ and $k' = M - (|k - \eta| \mod M)$ otherwise. Thus, by shifting in time we are delaying or advancing the time-stamp of interactions while by shifting in relations we are transferring the information of one relation $e_k$ into another one $e_k'$. Interestingly, this may provide a useful approach to model the flow of information in link streams. For example, suppose that node $u$ transmits a message to node $v$ at time $t$ and that there is a link stream $\mathcal{L}$ that is only registering this interaction. Namely,  $\mathcal{L}(t, e_k) = 1$ if $e_k = (u,v)$ and zero otherwise. Now, suppose that node $v$, after processing the message, sends a message to node $w$ at some future time $t + \tau$. Interestingly, if $(v,w)$ has index $k + \eta$ in $R$, then we have that the link stream representing this new interaction corresponds to $\mathcal{S}_{\tau, \eta}( \mathcal{L} )$. Hence, the flow of information corresponds to a time and relational shifting of a seed interaction. This seems particularly promising as a modeling technique for the case in which $R$ represents a path. Also, notice that (\ref{eq.shift}) reduces to the classical time series definition when $\mathcal{L}$ has no structure, as it not possible to move across relations in such case.

Equation (\ref{eq.correlation}) defined the zero-lag correlation between link streams. Based on our definition of shift, we can extend it to define the full cross-correlation function. In signal processing, the cross-correlation function is used to spot if two time series are similar even if they are out of phase. This is done by measuring the correlation between the two signals when one of them is shifted for all possible time lags. We thus define this concept for link streams as follows. Let $\mathcal{L}_1$ and $\mathcal{L}_2$ be two link streams defined on the same $T$ and $R$. The cross-correlation function between  $\mathcal{L}_1$ and $\mathcal{L}_2$ is defined as:
\begin{equation}\label{eq.cross_correlation}
(\mathcal{L}_1 \star \mathcal{L}_2)(\tau, \eta) = \text{corr}( \mathcal{L}_1, \mathcal{S}_{\tau, \eta}(\mathcal{L}_2) ) ~~\forall~\tau \in \mathbb{R}, \eta \in \mathbb{Z}. 
\end{equation}
The cross-correlation function can be useful to identify periodicities within a link stream. For example, suppose that $R$ represents the outgoing links of a router and that $\mathcal{I}$ ranks them in decreasing order of their bandwidth. It is normal that a router sends traffic via its large bandwidth links, albeit at peak hours it may rather do it via the small bandwidth ones in order to avoid congestions. Interestingly, if $\mathcal{L}$ models such traffic, then its redirection at peak times can be modeled as the shifting $\mathcal{L}(t_i + \tau', e_{k + \eta'}) = \mathcal{L}(t_i, e_k)$, where $(t_i, e_k)$ denotes the high bandwidth link at a normal time and $(t_i + \tau', e_{k + \eta'})$ the small bandwidth one at a congestion time. As a result, $(\mathcal{L} \star \mathcal{L})(\tau, \eta)$ peaks when $\tau = \tau'$ and $\eta = \eta'$, thus revealing the congestion times and redirection of traffic. Moreover, if such event occurs periodically, then $(\mathcal{L} \star \mathcal{L})(\tau, \eta)$ also peaks at multiples of $\tau$. Since correlation and shifting reduce to the time series definitions when applied on link streams with no structure, then (\ref{eq.cross_correlation}) does it too. 

{\bf Differentiation, gradient, Laplacian.}~Differentiation is another key concept in signal processing, permitting to measure the local variability of a time series. We extend this concept to link streams by measuring their local variability across the time and relational dimensions. Namely, let $\mathcal{L}$ be a link stream. We define its temporal differentiation as:
\begin{equation}\label{eq.time_derivative}
\frac{\partial \mathcal{L}}{\partial t} =\lim_{\tau \to 0} \frac{ \mathcal{S}_{- \tau, 0}(\mathcal{L}) - \mathcal{L} }{\tau} 
\end{equation}
Similarly, we define the relational differentiation of $\mathcal{L}$ as: 
\begin{equation}\label{eq.edge_derivative}
\frac{\partial \mathcal{L}}{\partial e} = \mathcal{S}_{0, -1}(\mathcal{L}) - \mathcal{L} 
\end{equation}
Notice that (\ref{eq.time_derivative}) and (\ref{eq.edge_derivative}) are link streams, therefore we can combine them into a time-relational derivative as:
\begin{equation}\label{eq.time_edge_derivative}
\frac{\partial^2 \mathcal{L}}{\partial e \partial t} = \frac{\partial}{\partial e}\left( \frac{\partial \mathcal{L}}{\partial t} \right) = \frac{\partial}{\partial t}\left( \frac{\partial \mathcal{L}}{\partial e} \right) 
\end{equation}
Equation (\ref{eq.time_edge_derivative}) can be interpreted as a measure of the extent to which similarly ranked edges in a link stream evolve similarly. For example, by retaking the case in which the link stream represents traffic in the internet backbone whose links have been ordered according to their bandwidth, we have that (\ref{eq.time_edge_derivative}) applied to it would measure (i) if links having similar bandwidths are handling similar amounts of traffic; and (ii) if the traffic in those links is evolving similarly.

In signal processing, when a signal is supported in more than one dimension (e.g., an image) its directional derivatives are stacked into a gradient vector. This vector is then leveraged to define a measure of the total variation of the signal and its Laplacian. We extend these concepts as follows. Given a link stream $\mathcal{L}$, we define its gradient vector as: 
\begin{equation}
(\nabla \mathcal{L})(t, e) = \left[\frac{\partial \mathcal{L}}{\partial t}(t,e), \frac{\partial \mathcal{L}}{\partial e}(t,e) \right]
\end{equation} 
The gradient therefore is a vector function supported on $T \times R$ containing the point-wise variation of $\mathcal{L}$ in each direction. As a result, if we compute the norm of such vector function (point-wise),  we obtain a link stream summarizing the (point-wise) variation of $\mathcal{L}$ in both directions. Let $ \| \nabla \mathcal{L} \|$ refer to the point-wise vector norm of the gradient. We have that the aggregation of all values in $\| \nabla \mathcal{L} \|$ produces a measure of the total variation of $\mathcal{L}$. Thus, we define the total variation of $\mathcal{L}$ as: 
\begin{equation}\label{eq.total_variation}
\text{TV}(\mathcal{L}) = \text{energy}(  \| \nabla \mathcal{L} \| )
\end{equation}
The Laplacian of a function is defined as the divergence of its gradient. We thus define the Laplacian of a link stream as the divergence of its gradient, given as:
\begin{equation}\label{eq.laplacian}
\Delta \mathcal{L} = \frac{\partial^2 \mathcal{L}}{\partial t^2} + \frac{\partial^2 \mathcal{L}}{\partial e^2}
\end{equation}
Notice that $\frac{\partial \mathcal{L}}{\partial e}(t,e) = 0$ for all $t$ when $R$ is a singleton. Therefore, for link streams with no structure, (\ref{eq.total_variation}) reduces to the classical Sobolev prior-based total variation measure used in time series, which also coincides with the definition of Dirichlet energy functional. Similarly, the Laplacian definition in (\ref{eq.laplacian}) trivially reduces to the classical one for time series when the link stream has no structure. 

\subsection{A Fourier-Structure transform for link streams} 
The Fourier decomposition is a technique that permits to express an arbitrary time series as a weighted combination of sine and cosine waveforms oscillating at different frequencies. The weighting coefficients, known as the Fourier transform, reveal the importance of each frequency in the time series. Our goal here is to develop an extension of the Fourier decomposition to link streams. This is, we aim to identify a set of elementary link streams that allow us to express a link stream as the weighted combination of them. Notice that we can readily apply the Fourier decomposition to a link stream $\mathcal{L}$ (with $T$ extended to $\mathbb{R}$) relation-wise as: 
\begin{equation}\label{eq.Fourier_edge_decomposition}
\mathcal{L}(t_i,e_k) = \int_{\mathbb{R}} c_f(e_k) \psi_{f}(t_i) df,
\end{equation}
where $\psi_{f}(t_i) = \exp\{j 2 \pi t_i f\}$, $j$ is the unit imaginary number,
\begin{equation}
c_f(e_k) = \int_{\mathbb{R}} \mathcal{L}(t, e_k) \psi^*_{f}(t) dt
\end{equation}
denotes the Fourier transform coefficient associated to frequency $f$ for relation $e_k \in R$, and $\psi^*$ refers to the complex conjugate of $\psi$. Since the waveforms $\psi_{f}$ are independent of $\mathcal{L}$, equation (\ref{eq.Fourier_edge_decomposition}) allows us to represent all the information of $\mathcal{L}$ through the coefficients $c_f(e_k)$. These coefficients summarize temporal information but not relational one, reason for which $c_f$, for a fixed $f$, remains a function of $e_k$. Therefore, to extend the Fourier transform to a relational dimension, we must make the coefficients aware of such information. 

We address this problem by developping an analog of the Fourier transform for the relational domain. This is, a decomposition for graphs that we can apply time-wise to the link stream in order to represent it through coefficients that now summarize relational information as a function of time. Then, we combine this graph decomposition with the Fourier one in order to develop a Fourier-Structure decomposition for link streams. To develop our decomposition for graphs, we notice that the Fourier transform fixes a set of very simple time series, the $\psi_k$'s, so that any other time series can be expressed as their combination. Thus, our graph decomposition follows the same approach: it fixes a set of very simple graphs so that, when properly combined, any other graph can be generated. To build this set, we may consider fixing a set of simple sub-structures such as cliques, triangles, stars, paths, etc., and then look for ways to combine them in order to generate other graphs. However, the right sub-structures to use seems highly application dependent. For instance, it may be more relevant to use cliques to represent interactions in a social network than paths. Yet, when considering traffic in a transportation network, the paths may be a better option. Thus, we develop a decomposition for graphs so that there is flexibility in the choice of elementary sub-structures to use. 

Let $D = \{R_i \subseteq R \}_{0 \leq i \leq p}$ denote the set of sub-structures that we aim to employ to decompose graphs, whose edge-sets are also a subset of $R$. We impose two conditions on these substructures: (i) $R_i \cap R_j = \emptyset$ for $i \neq j$; and (ii) $R = \cup_i R_i$. Therefore, we admit any dictionary of sub-structures as long as it forms a disjoint partitioning of the relational space. Yet, this great flexibility comes at the price that we must artificially enlarge $R$ by injecting virtual relations into each $R_i$ until we make the size of $R_i$ a power of two. This step is necessary in order to ensure the mathematical soundness of the decomposition, albeit we stress that such virtual additions do not pose any practical problem as the graphs to decompose will never contain them in their edge sets. Based on the enlarged sub-structures, let $\phi_i : R \to \mathbb{R}$ denote the normalized indicator function of $R_i$, i.e., $\phi_i(e) = 1/\sqrt{|R_i|}$ if $e \in R_i$ and zero otherwise. The set $D_{\phi} = \{\phi_i \}_{0 \leq i \leq p}$ therefore refers to the dictionary of indicator functions. 

Let $G_{R} = (V, E_{R}, g_{R})$ be any weighted subgraph of $R$, where $E_R \subseteq R$ refers to its edge-set and $g_R : R \to \mathbb{R}$ to its weight function. Our goal is therefore to express $g_R$ as a weighted combination of the $\phi_i$'s. We aim the weighting coefficient associated to $\phi_i$ to represent the importance of $R_i$ in $G_R$. A natural way to measure such importance is by counting the total weight that the relations of $R_i$ have in $G_R$. Thus, we propose to use $s_i = \frac{1}{\sqrt{|R_i|}}\sum_{e \in R_i} g_R(e)$ as the weighting coefficient of $\phi_i$. This allows us to express:
\begin{equation}\label{eq.graph_error_term}
g_R = \sum_{i} s_i \phi_i + \text{error}
\end{equation} 
The inclusion of the error term is necessary as there is no guarantee that our dictionary of sub-structures can represent every possible $g_R$. Take for instance the case in which we select a dictionary containing the complete graph as its only element $D = \{R_0 = R = V \times V$\}. Consider also that the graph $G_R$ that we aim to decompose contains only one edge. Then, clearly there is no weighting coefficient that can make a $\phi_0$ consisting of a complete graph become a $g_R$ containing only one non-zero entry. Thus, the error term in (\ref{eq.graph_error_term}) is included as a means to capture the information that cannot be represented through our dictionary of sub-structures. In the following, we will show that this error term can also be expressed through a dictionary whose elements indeed complete $D_{\phi}$ to become an orthonormal basis. Yet, we have less control over such elements, which consist of many sub-structures at different scales. Thus, it is important that the dictionary $D$ is initially well chosen to represent the data at hand in order to minimize the influence of the error term. 

Notice that $s_i \phi_i(e)$ is only non-zero for $e  \in R_i$ and that there is no other $s_k \phi_k(e)$ that is non-zero for $e \in R_i$. Therefore, the values of $g_R(e)$ for $e \in R_i$ are only being approximated by the term $s_i \phi_i$ in (\ref{eq.graph_error_term}). Moreover, observe that $s_i \phi_i(e) = \sum_{e' \in R_i} g_R(e') / |R_i|$ if $e \in R_i$ and zero otherwise. This means that $s_i \phi_i$ is approximating the values of $g_R(e)$ for $e \in R_i$ through their mean value. This shows that finding the error term that completes $s_i\phi_i$ amounts to solving the problem of recovering a set of values from their mean. Notably, this can be achieved by computing differences between the set of values at multiple resolutions. Namely, assume that we have $\{a,b,c,d\}$ and that we aim to recover them from their mean $s = (a+b+c+d) / 4$. To do it, we split the set as $\{a,b\}$ and $\{c,d\}$ and use the split to compute the mean differences: $w^{(0)} = ((a+b)-(c+d))/4$, $w^{(1)} = (a-b)/2$, $w^{(2)} = (c-d)/2$. Then, it is easy to check that $a = s + w^{(0)} + w^{(1)}$; $b = s + w^{(0)} - w^{(1)}$; $c = s - w^{(0)} + w^{(2)}$; and $d = s - w^{(0)} - w^{(2)}$. 

Thus, inspired from the procedure above, we set $R_i^{(0)} = R_i$ and recursively split this set in halves so that $R_i^{(\ell)} = R_i^{(2 \ell +1)} \cup R_i^{(2 \ell+2)}$. Then, we leverage this partitioning to define the functions $\theta_i^{(\ell)} : R \to \mathbb{R}$ given as: 
\begin{equation}
\theta_i^{(\ell)}(e) = 
\begin{cases}
        \frac{1}{\sqrt{|R_i^{(\ell)}|}} & \text{if } e \in R_i^{(2\ell + 1)} \\ \vspace{4pt}
        \frac{-1}{\sqrt{|R_i^{(\ell)}|}} & \text{if } e \in R_i^{(2\ell + 2)} \\ \vspace{4pt}
        0 & \text{otherwise }
    \end{cases}
\end{equation}
These functions therefore complete the dictionary $D_{\phi}$ and it is not hard to show that they turn it into an orthonormal basis for subgraphs of $R$. We thus project the graph $g_R$ into these elements in order to compute their weighting coefficient as $w_i^{(\ell)} = \sum_{e \in R} g_R(e) \theta_i^{(\ell)}(e)$, allowing us to replace the error term in (\ref{eq.graph_error_term}) as:
\begin{equation}\label{eq.graph_complete}
g_R = \sum_{i} s_i \phi_i + \sum_{i} \sum_{k} w_i^{(\ell)} \theta_i^{(\ell)}.
\end{equation}
Finally, given a link stream $\mathcal{L}$, we apply our relational decomposition time-wise. This allows us to express it as:
\begin{equation}
\mathcal{L}(t, e) = \sum_{i} s_i(t) \phi_i(e) + \sum_{i} \sum_{\ell} w_i^{(\ell)}(t) \theta_i^{(\ell)}(e),
\end{equation}
where $s_i(t)$ and $w_i^{(\ell)}(t)$ are computed as above but for $\mathcal{L}(t, e)$ with $t$ fixed and $e$ running. It remains to observe that $s_i(t)$ and $w_i^{(\ell)}(t)$ are time series that can be decomposed under the Fourier dictionary. Therefore, by decomposing them via Fourier, we obtain the Fourier-Structure decomposition of $\mathcal{L}$ as: 
\begin{multline}\label{eq.freq_struct_decomp}
\mathcal{L}(t, e) = \sum_{i} \int_{\mathbb{R}} s_{f,i} \left(\psi_f(t) \phi_i(e)\right) df \\ + \sum_{i} \sum_{\ell} \int_{\mathbb{R}} w_{f,i}^{(\ell)} \left(\psi_f(t) \theta_i^{(\ell)}(e)\right) df
\end{multline}
Interestingly, the products $\psi_f(t) \phi_i(e)$ and $\psi_f(t) \theta_i^{(\ell)}(e)$ produce a different value for every different combination of $e$ and $t$. Therefore, they correspond to link streams. In particular, they consist of link streams made of one structure $\phi_i$ or $\theta_i^{(\ell)}$ oscillating at frequency $f$, and the coefficients $s_{f,i}$ and $w_{f, i}^{(\ell)}$ indicate the extent at which such structures are oscillating at that frequency in $\mathcal{L}$. 

\subsection{Processing link streams with filters} 
A popular application of signal processing is filtering, aiming to amplify or suppress specific frequency information of a time series. In this subsection, we extend this concept to link streams, aiming to amplify or remove their frequency and structural information. Particularly, we show that we can use classical filters from signal processing to filter the frequency information of a link stream. We then develop the analog of filters for graphs and combine them with the classical time series ones in order to process both the frequency and structural information.

Time series filters can be seen as black boxes that receive a time series and produce a new one. If such black boxes are linear and time invariant, then they can be entirely characterized by their response to the impulse function. This is, processing any time series via that filter is equivalent to convolving the time series with the impulse response of the filter. Interestingly, we can use any classical time series filter as a means to process the frequency information of a link stream. To see this, let $h$ denote the impulse response of the filter and let us apply it to a link stream $\mathcal{L}$ relation-wise through the convolution sum:
\begin{equation}\label{eq.convolution_filter}
\tilde{\mathcal{L}}(t, e) = \int_{\mathbb{R}} h(\tau) \mathcal{L}(t - \tau, e) d\tau,
\end{equation}
where $\tilde{\mathcal{L}}$ refers to the filtered link stream. By decomposing $\mathcal{L}$ as in (\ref{eq.freq_struct_decomp}), then we have that:
\begin{multline}
\tilde{\mathcal{L}}(t, e) = \sum_{i} \int_{\mathbb{R}} \int_{\mathbb{R}} h(\tau) s_{f,i} \psi_f(t - \tau) \phi_i(e) df  d\tau \\ 
+ \sum_{i, \ell} \int_{\mathbb{R}} \int_{\mathbb{R}} h(\tau) w^{(\ell)}_{f,i} \psi_f(t - \tau) \theta^{(\ell)}_i(e) df  d\tau
\end{multline}
By exploiting the crucial property that $\psi_f(t - \tau) = \psi_f(t) \psi_f^{*}(\tau)$, then we have that:
\begin{multline}\label{eq.freq_filters}
\tilde{\mathcal{L}}(t, e) = \sum_{i} \int_{\mathbb{R}} s_{f,i} \gamma_f \psi_f(t) \phi_i(e) df  \\ 
+ \sum_{i, \ell} \int_{\mathbb{R}} w^{(\ell)}_{f,i}  \gamma_f \psi_f(t) \theta^{(\ell)}_i(e) df
\end{multline}
where $\gamma_f = \int_{\mathbb{R}} h(\tau) \psi_f^{*}(\tau) d\tau$ refers to the classical Fourier transform coefficient of $h$ associated to frequency $f$. Thus, (\ref{eq.freq_filters}) shows that $\tilde{\mathcal{L}}$ is the result of scaling the coefficients $s_{f,i}$, $w^{(\ell)}_{f,i}$ by a factor $\gamma_f$, implying that we can use any signal processing filter $h$ to process the frequency information of $\mathcal{L}$. To fully process a link stream, it therefore remains to construct filters that allow to target the structural information.

We develop structural filters by noticing that time series ones, through the convolution sum (\ref{eq.convolution_filter}), essentially replace the value of the time series at time $t$ by a weighted combination of the own time series values, where the weighting coefficients are given by $h$. Therefore, we build upon this observation to define the analog of filters and the convolution sum to the relational axis. Namely, let $q: R \times R \to \mathbb{R}$ denote a weighting kernel. Then, we define the relational convolution of $\mathcal{L}$ with the kernel $q$ as:
\begin{equation}\label{eq.convolution_relations}
\tilde{\mathcal{L}}(t, e) = \sum_{e' \in R} \mathcal{L}(t, e') q(e', e)
\end{equation}
This is, (\ref{eq.convolution_relations}) replaces the value of $\mathcal{L}$ at point $(t,e)$ by a weighted average of all the other values at time $t$, where the weighting coefficients are given by $q$ and may be different for each relation. Let us now assume that we aim to filter $\mathcal{L}$ structure wise, so that its coefficients $s_{f,i}$ and $w^{(\ell)}_{f,i}$ are re-scaled by coefficients $\sigma_{i}$ and $\nu^{(\ell)}_{i}$, respectively. Then, we have that $q$ represents such a filter if it is constructed as:
\begin{equation}
q_{filt}(e', e) = \sum_{i} \sigma_i \phi_i(e') \phi_i(e) + \sum_{i \ell} \nu^{(\ell)}_{i} \theta^{(\ell)}_i(e') \theta^{(\ell)}_i(e)  
\end{equation}
To demonstrate it, we perform the relational convolution between $\mathcal{L}$ and $q_{filt}$ while decomposing $\mathcal{L}$:
\begin{multline}\label{eq.graph_filter_full}
\tilde{\mathcal{L}}(t, e) = \int_{\mathbb{R}} \sum_{i, j} \sum_{e'} c_{f,i} \sigma_j \psi_f(t) \phi_i(e') \phi_j(e') \phi_j(e) df  \\
+  \int_{\mathbb{R}} \sum_{i, j, \ell} \sum_{e'} w^{(\ell)}_{f,i} \nu^{(\ell)}_j \psi_f(t) \theta^{(\ell)}_i(e') \theta^{(\ell)}_j(e') \theta^{(\ell)}_j(e) df 
\end{multline}
Due to the orthonormality of the graph dictionary, we have that $\sum_{e'} \phi_i(e') \phi_j(e') = 1$ if $i = j$ and zero otherwise (same for $\theta_i^{(\ell)}$ and  $\theta_j^{(\ell)}$). Therefore, (\ref{eq.graph_filter_full}) reduces to:
\begin{multline}
\tilde{\mathcal{L}}(t, e) = \int_{\mathbb{R}} \sum_{i}  c_{f,i} \sigma_i \psi_f(t) \phi_i(e) df  \\
+  \int_{\mathbb{R}} \sum_{i, \ell} \sum_{e'}w^{(\ell)}_{f,i}  \nu^{(\ell)}_i \psi_f(t) \theta^{(\ell)}_i(e) df 
\end{multline} 
showing that $c_{f,i}$ and $w^{(\ell)}_{f,i}$ are indeed re-scaled by $\sigma_i$ and $\nu^{(\ell)}_i$, as desired. Clearly, (\ref{eq.convolution_filter}) and (\ref{eq.convolution_relations}) can be jointly applied to $\mathcal{L}$, thus permitting its joint frequency-structure filtering. We finish stressing that given that our decomposition and filters handle the temporal and relational dimensions separately, when $\mathcal{L}$ has no structure then they trivially reduce to the classical definitions for time series.

\section{Conclusion}\label{Sec.conclusion}

In this paper, we showed that link streams can be seen as a generalization of graphs and time series. While the graph extension naturally arises by studying the link stream as a set of time-stamped edges, the time series ones appears when the link stream is studied through the weight function of such time-stamped edges. In particular, we showed that this weight function corresponds to an extension of time-series to a relational dimension, for which we extended numerous concepts from signal processing to consider a relational dimension, in addition to the temporal one. Our extensions range from elementary concepts, like energy and differentiation, to more advanced ones like the Fourier transform and filters. We showed that some of these extensions show interesting connections with the formalism developed from the graph perspective when the link stream is unweighted. Thus, they offer a starting point to develop extensions of the graph concepts to a weighted scenario and perhaps to the convergence of the two visions. Some of the considered concepts depend on the ordered nature of time, thus we extended them by also assuming the relational axis to be ordered. While less intuitive, we highlighted that through simple orders of the relational axis some interesting analyses can be done with the proposed concepts. The next steps consist in assessing the pertinence of the proposed concepts in real-world applications, continue the extension of the graph and signal frameworks to link streams, and explore more in depth the connection between the two perspectives. 

\medskip
\noindent
{\bf Acknowledgements.}
This work is funded in part by the ANR (French National Agency of Research) through the ANR FiT LabCom.

\bibliographystyle{ieeetr}
\bibliography{biblio}
\end{document}